\documentclass[aps,prb,reprint,superscriptaddress]{revtex4-2}
\usepackage{graphicx}% Include figure files

\begin{document}

\title{Anisotropic magnetotransport properties of the heavy-fermion superconductor CeRh$_2$As$_2$}

\author{S.~Mishra}\email[]{sanu@lanl.gov}\affiliation{Los Alamos National Laboratory, Los Alamos, New Mexico, 87545, USA}

\author{Y. Liu}\affiliation{Los Alamos National Laboratory, Los Alamos, New Mexico, 87545, USA}

\author{E. D. Bauer}\affiliation{Los Alamos National Laboratory, Los Alamos, New Mexico, 87545, USA}

\author{F. Ronning}\affiliation{Los Alamos National Laboratory, Los Alamos, New Mexico, 87545, USA}

\author{S. M. Thomas}\affiliation{Los Alamos National Laboratory, Los Alamos, New Mexico, 87545, USA}

\date{\today}

\begin{abstract}
We report anisotropic resistivity measurements of the heavy-fermion superconductor CeRh$_2$As$_2$ in magnetic fields up to 16~T and temperatures down to 0.35~K. The measured CeRh$_2$As$_2$ resistivity shows a signature corresponding to the suggested quadrupole density wave order state at $T_0 \sim$ 0.5~K for both measured directions. For a magnetic field applied along the tetragonal $a$ axis, $T_0$ is enhanced with magnetic field reaching $\sim$1.75~K at 16~T. Further, a magnetic field-induced transition occurs at $\mu_0 H_m \sim $ 8.1~T corresponding to a change to a new broken symmetry state. For a magnetic field applied along the $c$ axis, $T_0$ is suppressed below our base temperature $\sim$0.35~K by $\mu_0 H \sim$ 4.5~T, a field close to the previously reported field-induced transition within the superconducting state suggested to be from an even-parity to an odd-parity state. Our results indicate that the multiple superconducting phases in CeRh$_2$As$_2$ are intimately tied to the suppression of the proposed quadrupole-density-wave phase at $T_0$.
\end{abstract}

\maketitle

\subsection{Introduction}
Systems exhibiting multiple superconducting phases are a rarity in nature. Only a handful of such compounds are known to exist, prominent among them are UPt$_3$ \cite{Joynt}, thorium-doped UBe$_{13}$ \cite{Heffner}, PrOs$_4$Sb$_{12}$ \cite{Bauer, Izawa, Maple} and UTe$_2$ \cite{Braithwaite2019,Aoki2020, Thomas}. The multiple superconducting phases in these systems are proposed to be useful in the pursuit of topological quantum computation. The recently discovered heavy-fermion superconductor CeRh$_2$As$_2$ is a recent addition to this exotic class of systems \cite{Khim}.

CeRh$_2$As$_2$ crystallizes in the CaBe$_2$Ge$_2$-type centrosymmetric tetragonal crystal structure (space group $P4/nmm$) \cite{Madar1987} as shown in Fig. \ref{fig:CeRh2As2structure}(a). It becomes superconducting (SC) below $T_c$ = 0.26~K. The upper critical magnetic fields along both the principal axes (i.e., $\mu_0 H \parallel a \approx$ 2~T and $\mu_0 H \parallel c \approx$ 14~T) well exceed the Pauli paramagnetic limit $\mu_0 H_{PPL} =$ 1.86(T/K)$T_c \sim$ 0.5~T. For a magnetic field applied along the $c$ axis, a field-induced transition occurs within the superconducting state at $\mu_0 H^* \sim$ 4~T, which has been suggested to correspond to a change of the superconducting state from an even-parity (SC1) to an odd-parity state (SC2). In contrast, for a field applied in the $ab$ plane only one superconducting phase exists \cite{Khim,Landaeta}. Interestingly, nuclear quadrupole resonance measurements suggest an antiferromagnetic order also exists within the superconducting state \cite{Kibune}.

Another intriguing second-order phase transition at $T_0 \sim$ 0.4~K was identified precursing the superconducting state and was suggested to be a nonmagnetic quadrupole-density-wave (QDW) phase (I) \cite{Hafner}. A feature corresponding to $T_0$ was observed in specific heat, thermal expansion and resistivity \cite{Khim, Hafner}. For a magnetic field applied in the $ab$ plane, a field-induced transition to a new phase was observed at $\mu_0 H_m \sim$ 9~T in magnetostriction, resistivity, magnetization, and magnetic torque \cite{Hafner}.

The origin of the multiple superconducting phases due to the field-induced transition at $\mu_0 H^* \sim$ 4~T in CeRh$_2$As$_2$ remains elusive. Currently, the unique crystal structure of CeRh$_2$As$_2$ is believed to lie at the heart of its unique superconducting properties. In CeRh$_2$As$_2$, although the global inversion symmetry is preserved, the local inversion symmetry is lacking as Ce atoms have different Rh and As environments above and below it. This enables a staggered Rashba spin-orbit coupling and is suggested to be responsible for the field-induced transition within the superconducting state \cite{Khim}. Thus, its nonsymmorphic crystal structure is suggested to play a key role in the field-induced transition \cite{Ptok, Cavanagh}. Khim $et al.$ \cite{Khim}, however, did not exclude the possibility of normal state properties affecting the superconducting properties in CeRh$_2$As$_2$. They pointed out that the feature in specific heat at $T_0$ also seemed to be suppressed by a field $\mu_0 H \sim$ 4~T. The specific heat feature, however, is subtle and therefore requires more careful studies.

Nonetheless, the rich phase diagram of CeRh$_2$As$_2$ raises the question of the relationship between the multiple phases, especially the superconducting state and the suggested QDW phase below $T_0$. In this regard, the unique behavior of the superconducting phase in applied magnetic fields along the two principal crystallographic directions, i.e., $a$ and $c$ axes warrants a similar anisotropic investigation of the suggested QDW phase below $T_0$. Furthermore, in the previous transport study \cite{Hafner}, the current was applied in the basal plane of the tetragonal structure based on which it was suggested that the propagation vector of the order parameter of the phase below $T_0$ has a component within the basal plane. To the best of our knowledge, transport studies with current applied perpendicular to the basal plane i.e., along the $c$ axis are still lacking. Therefore, an important piece of information regarding the order parameter of the $T_0$ phase remains missing. To remedy these shortcomings, we performed anisotropic magnetotransport measurements on microstructured devices made out of a single crystal of CeRh$_2$As$_2$.

In this Letter, we present anisotropic magnetotransport measurements in CeRh$_2$As$_2$. Our main findings are as follows : (1) CeRh$_2$As$_2$ is weakly anisotropic with the in-plane resistivity being roughly three-fourths the interlayer resistivity. (2) A signature corresponding to the suggested QDW phase at $T_0$ is also observed in the out-of-plane resistivity. (3) For a magnetic field applied along the $c$-axis, $T_0$ is suppressed below our base temperature $\sim$0.35~K by a magnetic field $\mu_0 H\sim$ 4.5~T, a field, possibly coincidentally, similar to the previously reported field-induced even-odd parity transition within the superconducting state.

\subsection{Experimental details}
\begin{figure}[h]
	\centering
    \includegraphics[width=\columnwidth]{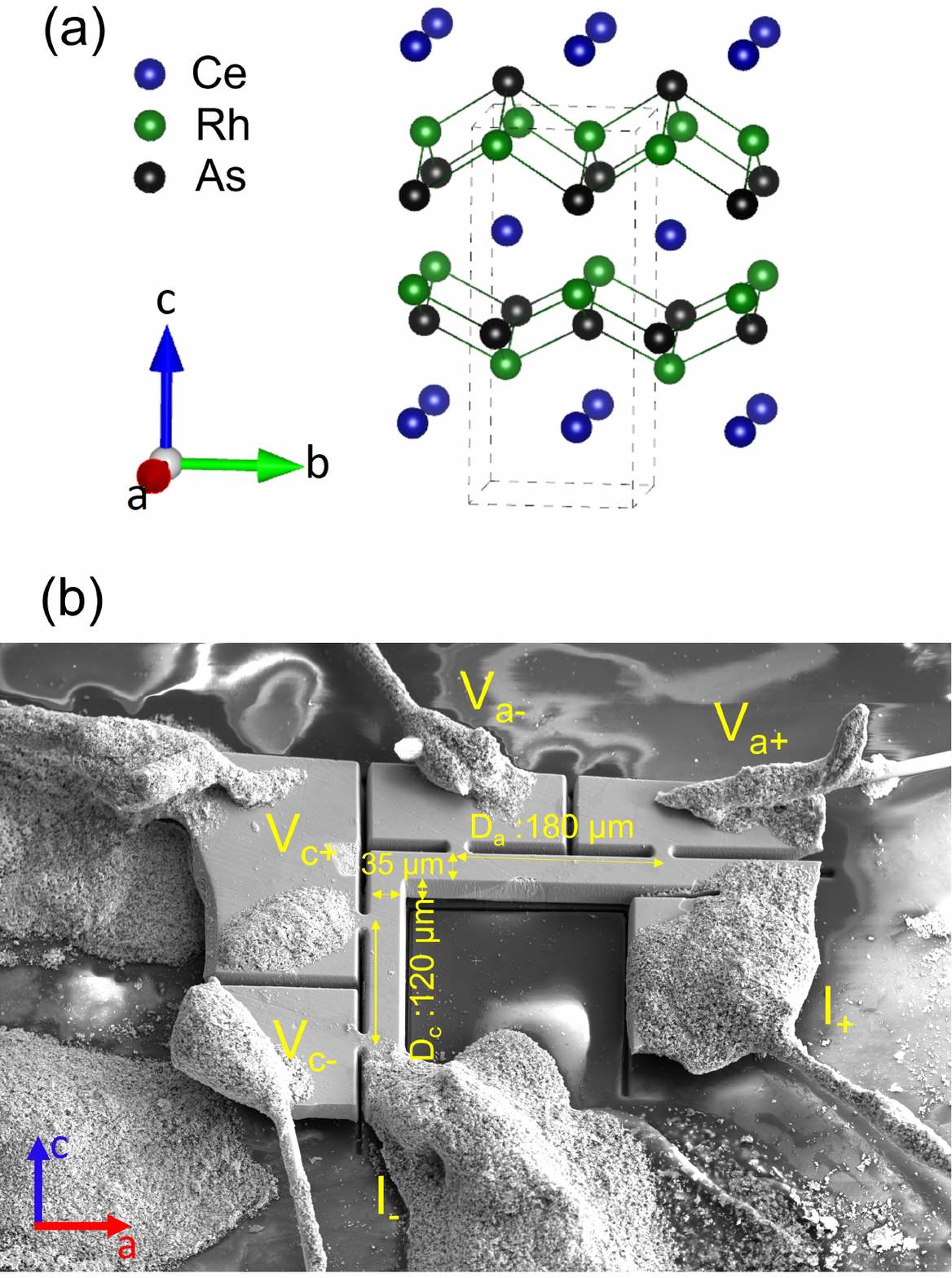}
\caption{\label{fig:CeRh2As2structure}(a) CaBe$_2$Ge$_2$-type tetragonal crystal structure of CeRh$_2$As$_2$ with the elements Ce, Rh, and As labelled in blue, green, and black, respectively. The tetragonal unit cell is depicted by the dashed box. (b) FIB-fabricated microstructured device on a CeRh$_2$As$_2$ single crystal. $D_a$ and $D_c$ represent the two bar-shaped sections of the device, aligned along the crystallographic $a$ and the $c$ axes, respectively. Current and voltage leads as well as the device dimensions are labeled.}

\end{figure}
Single crystals of CeRh$_2$As$_2$ were grown in Bi flux starting from a mixture of pure elements Ce, Rh, As, and Bi with a molar ratio of $1 : 2 : 2 : 30$. Starting materials were sealed in an evacuated fused silica tube, which was heated to 1150 $^\circ$C over 30~h, followed by a dwell at 1150 $^\circ$C for 24~h, then gradually cooled to 700 $^\circ$C at a rate of 2.5 $^\circ$C$/$h. The crystallographic structure of CeRh$_2$As$_2$ was verified at room temperature by a Bruker D8 Venture single-crystal x-ray diffractometer equipped with Mo radiation. X-ray diffraction analysis shows that CeRh$_2$As$_2$ crystallizes in the tetragonal space group $P4/nmm$ (No. 129) with lattice parameters $a = b =$ 4.28~\r{A} and $c =$ 9.85~\r{A}, in agreement with a previous report \cite{Khim}. The transport measurements were performed using a Quantum Design physical property measurement system (QDPPMS) equipped with a $^3$He option reaching a base temperature of $\sim$ 350~mK. For zero-field measurements an adiabatic demagnetization refrigerator extended the base temperature down to $\sim$ 100~mK. The four-wire resistance was measured using an ac resistance bridge (Lakeshore model 372). The anisotropic-transport measurements were performed on a microstructured device of CeRh$_2$As$_2$ fabricated using focused ion-beam (FIB) milling from a single crystal as shown in Fig. \ref{fig:CeRh2As2structure}(b). After Laue-orienting the single crystal, it was polished along the $ac$-plane down to a thickness of 35 $\mu$m. Out of the oriented single crystal, a six-terminal L-shaped microstructured device was fabricated to measure the anisotropic electrical transport properties along the two bar-shaped sections aligned with crystallographic $a$ ($D_a$) and $c$ ($D_c$) axes as shown in Fig. \ref{fig:CeRh2As2structure}(b). The dimensions ($l \times w \times h$) of the two bar-shaped sections with current along $a$ and $c$ axes, namely D$_a$ and D$_c$, are 180 $\mu m$ $\times$ 35 $\mu m$ $\times$ 35 $\mu m$ and 120 $\mu m$ $\times$ 35 $\mu m$ $\times$ 35 $\mu m$, respectively. The current and voltage leads for the device are labelled in Fig. \ref{fig:CeRh2As2structure} (b).

\subsection{Results}
\begin{figure}
	\centering
    \includegraphics[width=\columnwidth]{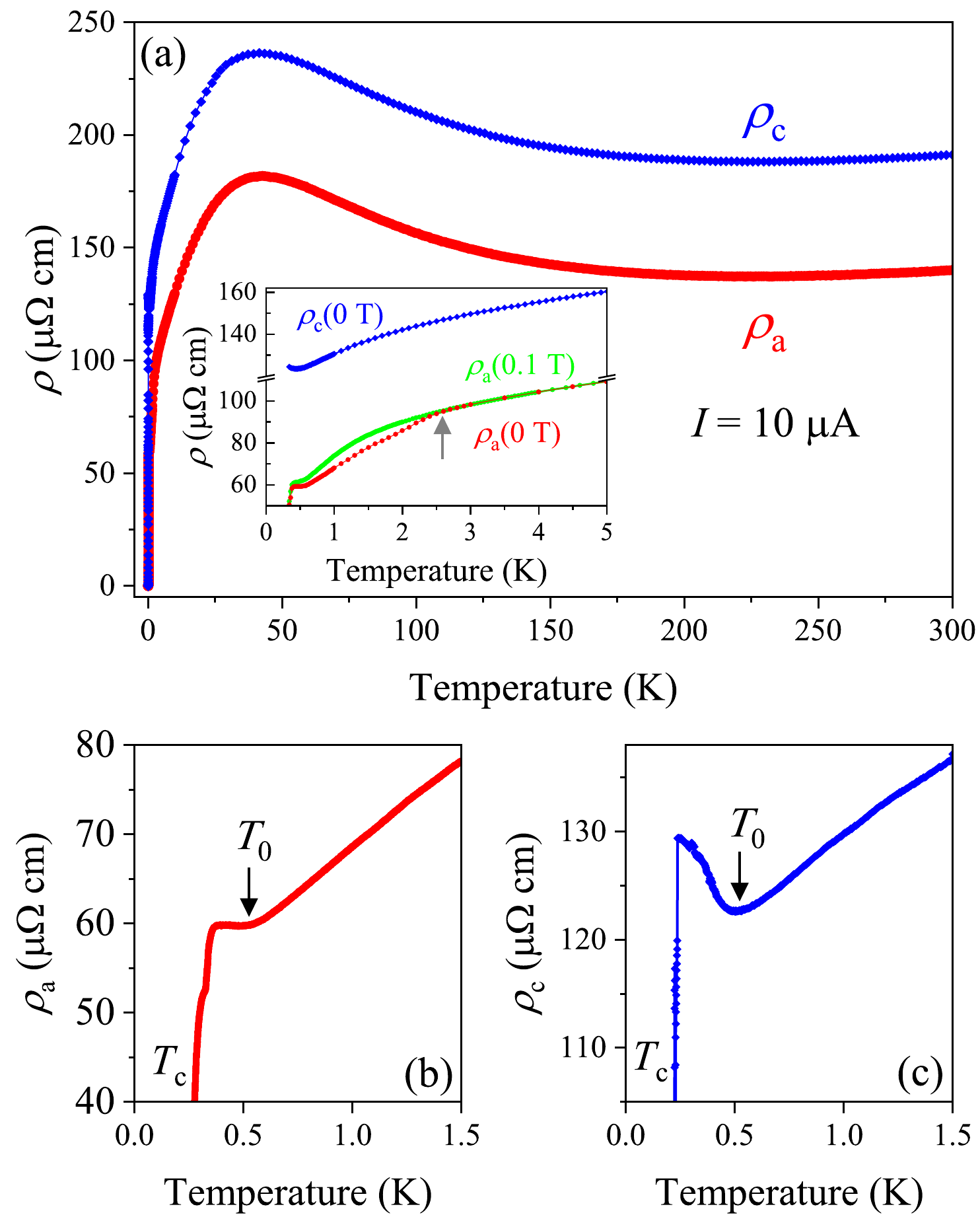}
\caption{\label{fig:RvsT} (a) Resistivities $\rho_a$ and $\rho_c$ for a current $I =$ 10 $\mu $A applied to the microstructured sections $D_a$ and $D_c$, respectively. Inset shows the resistivities $\rho_a$(0~T),  $\rho_a$(0.1~T) and $\rho_c$(0~T). Panel (b) and (c) shows a zoomed in view of the low-temperature resistivities $\rho_a$ and $\rho_c$, respectively.}
\end{figure}

Figure \ref{fig:RvsT}(a) shows the anisotropic resistivities $\rho_a$($T$) and $\rho_c$($T$) of CeRh$_2$As$_2$ for a current $I =$ 10 $\mu $A applied in the microstructured device along the two bar-shaped sections $D_a$ and $D_c$ i.e., along the $a$ and the $c$ axes of the tetragonal system, respectively. The superconducting transition, marked by a drop to the zero resistivity state, occurs at $T_c \sim$ 0.26~K, in agreement with that reported through bulk thermodynamic probes \cite{Khim,Landaeta,Hafner}. The superconducting transition is sharper in $\rho_c$($T$) compared to $\rho_a$($T$). CeRh$_2$As$_2$ shows the typical resistive behaviour of a heavy-fermion system. The broad humplike feature in both $\rho_a$($T$) and $\rho_c$($T$) around $\sim$ 40~K corresponds to the development of a coherent Kondo lattice. As evident from Fig \ref{fig:RvsT}(a), the in-plane resistivity ($\rho_a$) is roughly three-fourths the inter-layer resistivity ($\rho_c$), implying that the transport properties of CeRh$_2$As$_2$ are weakly anisotropic.

The inset in Fig. \ref{fig:RvsT}(a) shows resistivities $\rho_a$($T$) and $\rho_c$($T$) measured at zero-field as well as $\rho_a$($T$) measured at $\mu_0 H =$ 0.1 T. The gray arrow points to the resistance drop at $\sim$2.5~K due to the inclusion of a bismuth-rich superconducting impurity as determined by energy-dispersive x-ray spectroscopy in the $D_a$ bar. This resistance drop vanishes in a field as small as 0.1 T. The impurity can be seen as a light patch in scanning electron microscopy (SEM) image of Fig. \ref{fig:CeRh2As2structure}(b). No such impurity inclusion exists in the $D_c$ bar.

Below the coherence temperature, $\rho_a$($T$) and $\rho_c$($T$) decrease rapidly down to $T_0 \sim$ 0.5~K. Below $T_0$, $\rho_a$($T$) shows little temperature-dependence down to the superconducting transition. In contrast, $\rho_c$($T$) exhibits a strong upturn and continues to increase down to the superconducting transition [see Figs. \ref{fig:RvsT}(b) and \ref{fig:RvsT}(c)]. For $\rho_a$($T$), we define $T_0$ as the temperature where $\partial ^2 \rho_a (T)/ \partial T^2$ has a local extremum as marked by an arrow in Fig. \ref{fig:RvsT}(b). In $\rho_c$($T$), this corresponds to the local minima before the strong upturn and therefore we take the local minimum as $T_0$ for $\rho_c$($T$), as marked by arrow in Fig. \ref{fig:RvsT}(c).

Previously, a suggestion of a gap opening at the Fermi level was made based on the behavior of $\rho_a$($T$) below $T_0$ \cite{Hafner}. This signature of $T_0$ in $\rho_a$($T$), in conjunction with its behavior observed in other measurements, was suggested to be an indication of a QDW state \cite{Hafner}. Similarly, our observation of the strong increase in $\rho_c$($T$) below $T_0$ suggests the opening of a gap at the Fermi level. Furthermore, the presence of a strong upturn in $\rho_c$($T$) in comparison to $\rho_a$($T$) at $T_0$ suggests that the propagation vector of the order parameter of the phase below $T_0$ also has an out-of-plane component, in addition to an in-plane component reported previously \cite{Hafner}.

\begin{figure}
	\centering
    \includegraphics[width=\columnwidth]{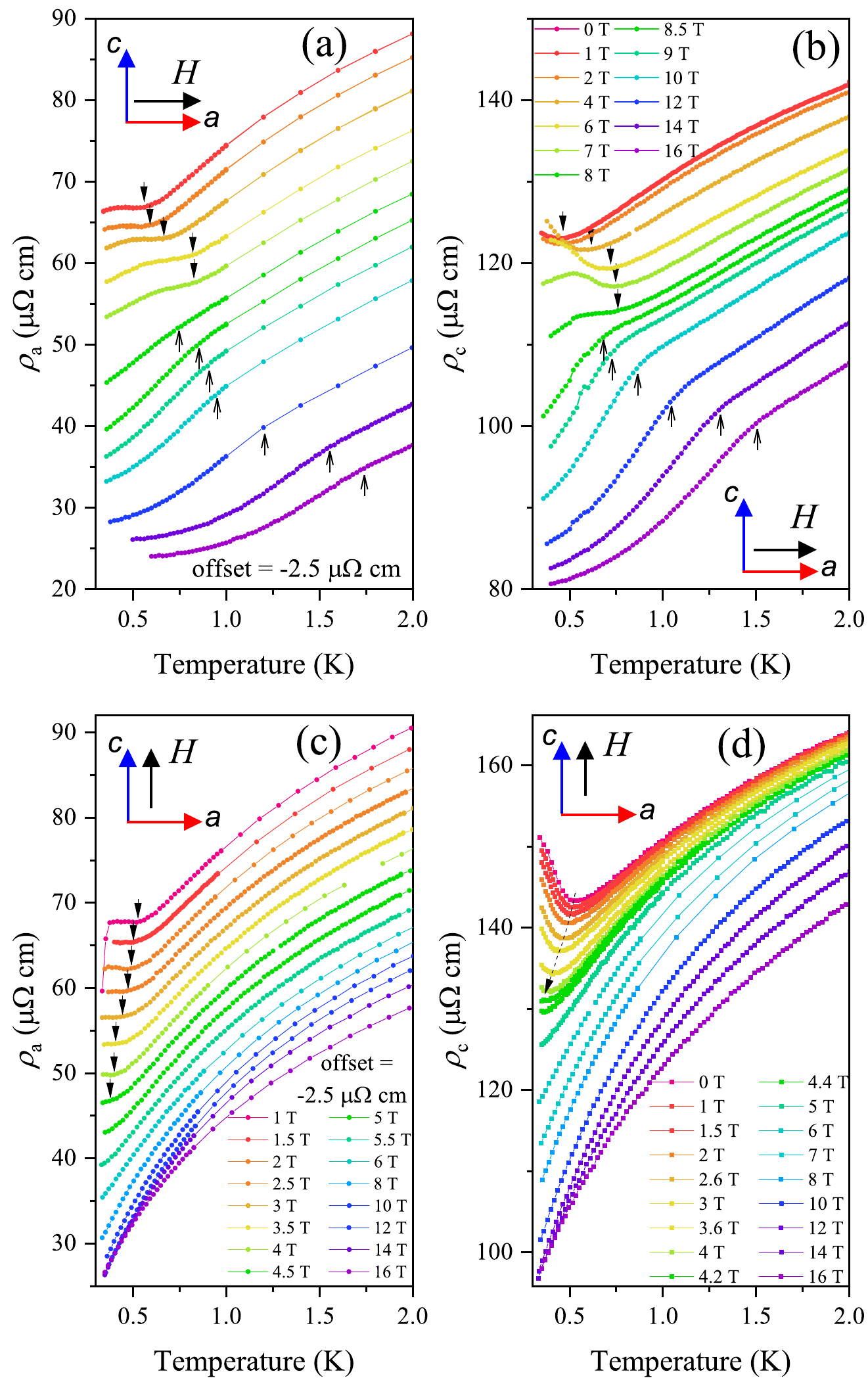}
\caption{\label{fig:RvsTatallB} (a) $\rho_a (T)$ and (b) $\rho_c (T)$ at different constant fields applied along the $a$ axis of CeRh$_2$As$_2$. Solid (open) black arrows point to the $T_0$ feature below (above) 8~T. (c) and (d) show the corresponding plots for field applied along the $c$ axis of CeRh$_2$As$_2$. Black arrows point to the feature at $T_0$. In (a) and (c), the curves are shifted vertically for clarity. The legend for (a) is the same as (b).}
\end{figure}

Next, to determine the field evolution of the phase below $T_0$, we measured $\rho_a (T)$ and $\rho_c (T)$ at several constant magnetic fields up to 16~T applied along both the principal axes of the tetragonal system, i.e., $a$ and $c$ axes as shown in Fig. \ref{fig:RvsTatallB}. The open and solid black arrows point to the feature at $T_0$ at different fields.
\begin{figure}
	\centering
    \includegraphics[width=\columnwidth]{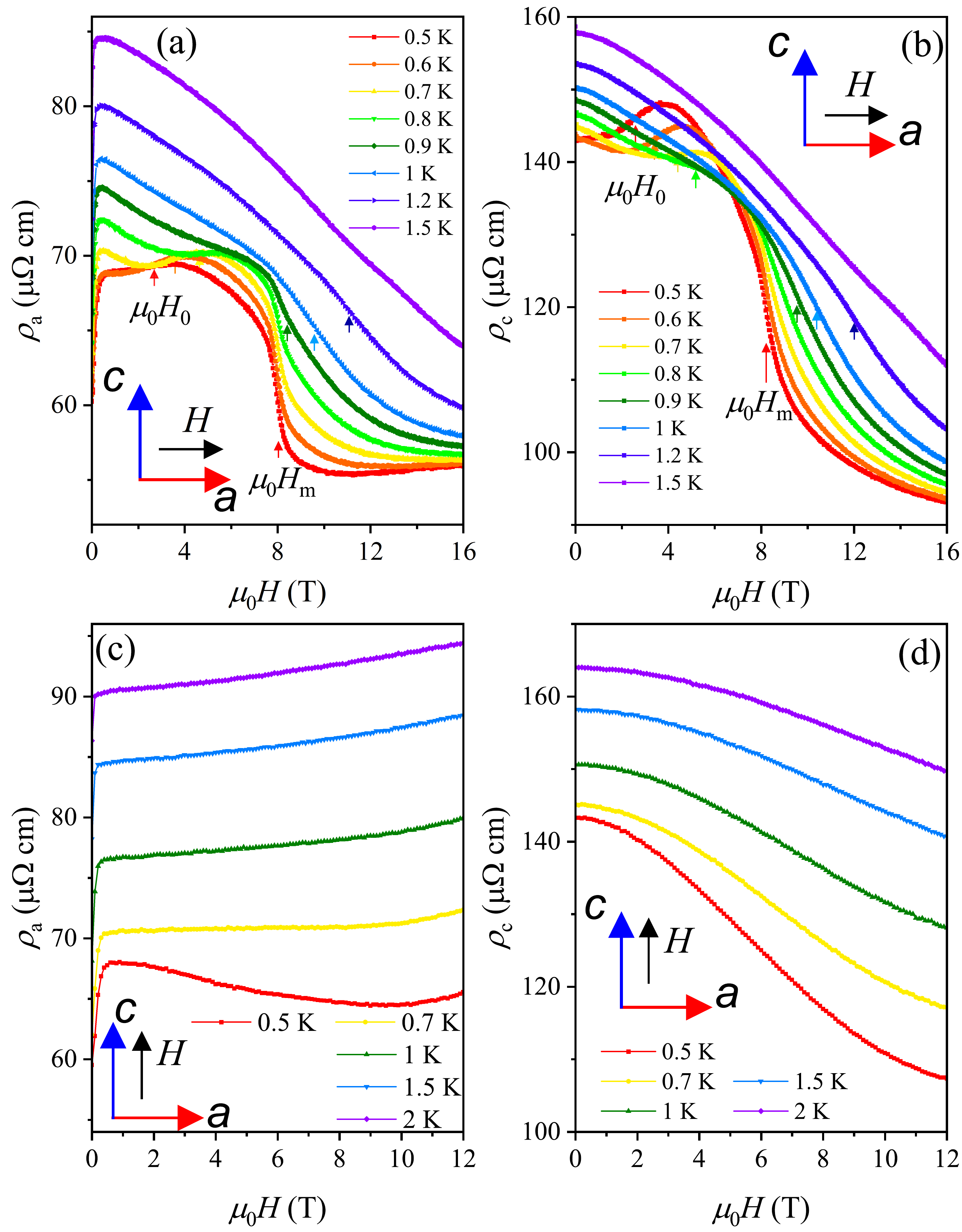}
\caption{\label{fig:Magnetoresistance1} (a) $\rho_a$($\mu_0 H$) and (b) $\rho_c$($\mu_0 H$) at several constant low-temperatures for a field applied along the $a$-axis of CeRh$_2$As$_2$. The field-induced transitions $\mu_0 H_m$ and $\mu_0 H_0$ are marked by arrows. (c) $\rho_a$($\mu_0 H$) and (d) $\rho_c$($\mu_0 H$) at several constant low-temperatures for a field applied along the $c$-axis of CeRh$_2$As$_2$.}
\end{figure}
It is clearly evident from Fig. \ref{fig:RvsTatallB} that $T_0$ evolves differently for the two field orientations. For a same field orientation, i.e., $H \parallel a$ or $H \parallel c$, $T_0$ evolves similarly in both $\rho_a (T)$ and $\rho_c (T)$. For $H \parallel a$, a clear distinction is evident in both $\rho_a (T)$ and $\rho_c (T)$ at fields above and below 8~T as marked by solid black ($\mu_0 H \le$ 8~T) and open black ($\mu_0 H \ge$ 8~T) arrows in Figs. \ref{fig:RvsTatallB}(a) and \ref{fig:RvsTatallB}(b). The subtle plateau in $\rho_a (T)$ and the upturn in $\rho_c (T)$ vanish near 8~T and becomes a downturn consistent with a transition into a new phase above 8~T. Therefore, above 8~T, $T_0$ is obtained from the local minimum in $\partial^2 \rho_c(T)/ \partial T^2$ and $\partial^2 \rho_c(T)/ \partial T^2$. In contrast, for $H \parallel c$, $T_0$ is monotonously suppressed with field and there is no indication of a phase transition down to 0.35~K for $\mu_0 H \ge$ 4.5~T, as shown in Figs. \ref{fig:RvsTatallB}(c) and \ref{fig:RvsTatallB}(d). Measurements to lower temperatures are necessary to uncover the fate of $T_0$ in relation to superconductivity.

To further elucidate the phase diagram, we measured the anisotropic magnetoresistances i.e., $\rho_a$($\mu_0 H$) and $\rho_c$($\mu_0 H$) at several constant low temperatures for fields applied along the $a$ and the $c$ axes of CeRh$_2$As$_2$ as shown in Fig. \ref{fig:Magnetoresistance1}.
\begin{figure}
	\centering
    \includegraphics[width=\columnwidth]{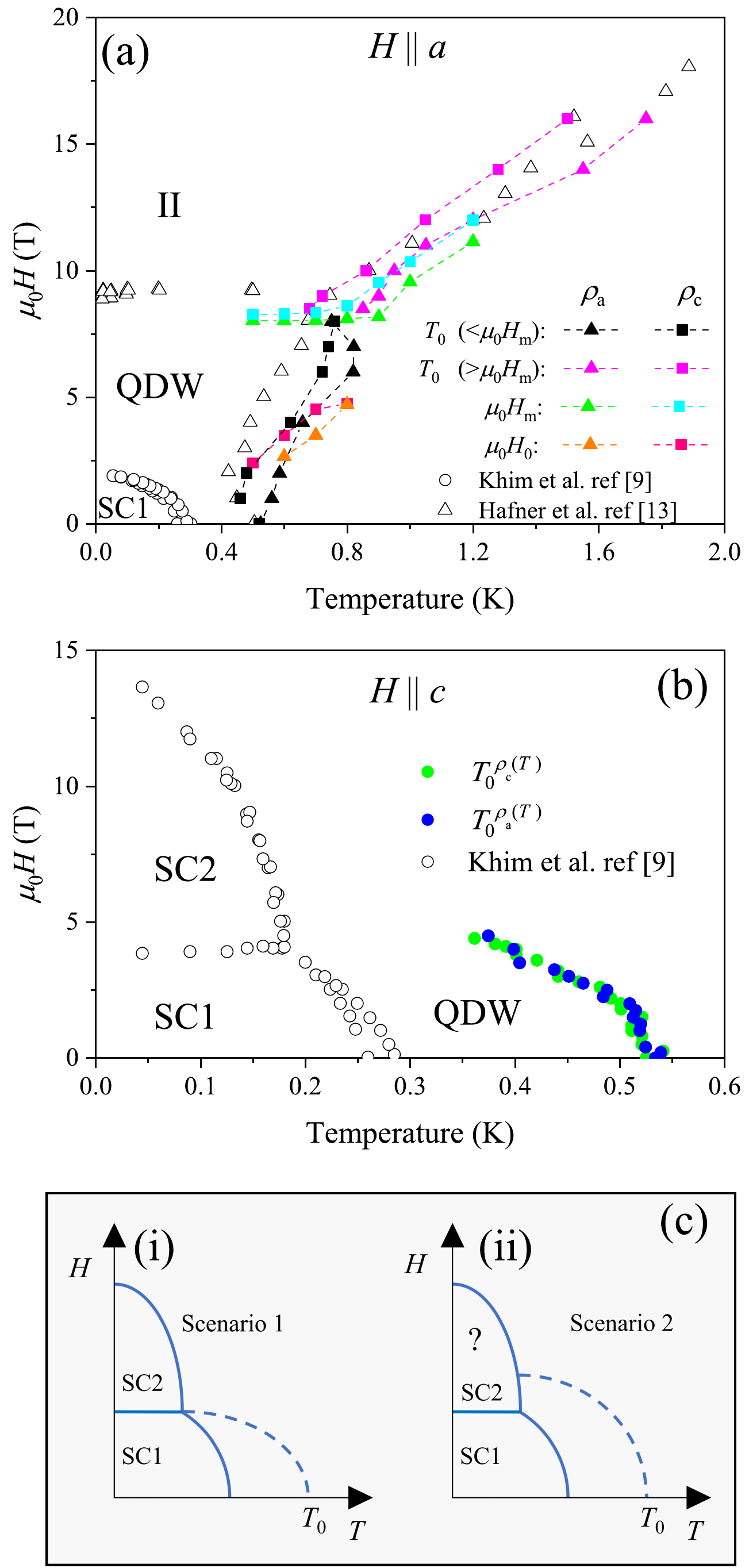}
\caption{\label{fig:Phasediagram1} Temperature magnetic field ($T-H$) phase diagram for CeRh$_2$As$_2$ based on our magnetotransport measurements for field applied along (a) the $a$ and (b) the $c$ axes. In (a) magenta (solid black) symbols correspond to $T_0$ features in $\rho (T)$ at field above (below) $\mu_0 H_m$. Solid triangles (squares) correspond to the features in $\rho_a$ ($\rho_c$). Green triangles (cyan squares) correspond to the feature at $\mu_0 H_m$ in $\rho_a$($\mu_0 H$) [$\rho_c$($\mu_0 H$)]. In (b) blue (green) solid circles correspond to the $T_0$ features in $\rho_a (T)$ [$\rho_c (T)$]. The data corresponding to the open circles are taken from Ref. \cite{Khim} and the open triangles is taken from Ref. \cite{Hafner}. (c) Two scenarios for the $T_0$ phase boundary as discussed in the text.}
\end{figure}
For field applied along the $a$ axis, there is a distinct field-induced transition occurring at $\mu_0 H_m \sim$ 8.1~T for temperatures below $\sim$1.2~K as shown in Figs. \ref{fig:Magnetoresistance1} (a) and \ref{fig:Magnetoresistance1}(b). The transition is sharper in $\rho_a$($\mu_0 H$) relative to $\rho_c$($\mu_0 H$) possibly suggesting the states that contribute more to the in-plane electronic properties are more strongly affected by the phase transition. For fields applied in the $ab$-plane, a field-induced transition was previously observed at $\sim$9~T and was suggested to correspond to a change from QDW (I) to another non-magnetic phase (II) \cite{Hafner}. In addition, there is another subtle feature in both $\rho_a$($\mu_0 H$) and $\rho_c$($\mu_0 H$) at $\mu_0 H_0 \sim$ 2.5~T at 500~mK. Its temperature evolution is marked by arrows in Fig. \ref{fig:Magnetoresistance1}(a) and (b), respectively.  On the other hand, for field applied along the $c$ axis, no such field-induced transitions are apparent in either $\rho_a$($\mu_0 H$) or $\rho_c$($\mu_0 H$). The sharp increase in $\rho_a$($\mu_0 H$) at very low fields for both $H \parallel a$ and $H \parallel c$ is extrinsic to CeRh$_2$As$_2$ and attributed to a superconducting impurity inclusion discussed previously.

\subsection{Discussion}
We plot a revised temperature magnetic field ($T-H$) phase diagram for CeRh$_2$As$_2$ based on Fig. \ref{fig:RvsTatallB} and \ref{fig:Magnetoresistance1}, along with data from Refs. \cite{Khim,Hafner}, as shown in Fig. \ref{fig:Phasediagram1}. Our phase diagram for $H \parallel a$ agrees well with Ref. \cite{Hafner}. The proposed QDW phase that exists below $T_0 \sim$ 0.5~K at zero field extends to higher temperatures with increasing fields. At the highest field of our measurement, i.e., 16~T, $T_0$ occurs at $\sim$ 1.75~K. Furthermore, a clear field-induced transition occurs at $\mu_0 H_m \approx$ 8.1~T corresponding to a change from the suggested QDW state to a new broken symmetry state below $T_0$. A second phase transition exists at $\mu_0 H_0$ corresponding to the $T_0$ feature in $\rho_a$($\mu_0 H$) and $\rho_c$($\mu_0 H$).

The most interesting result is obtained for $H \parallel c$. In both $\rho_a (T)$ and $\rho_c (T)$, the feature at $T_0$ is suppressed below our base temperature $\sim$0.35~K by a field $\mu_0 H \sim$ 4.5~T. This suppression field is close to the one for the field-induced transition from an even-parity (SC1) to odd-parity (SC2) superconducting state shown by the open circles in Fig. \ref{fig:Phasediagram1}(b) based on Ref. \cite{Khim} which may be a coincidence.

Based on this observation, we envision two scenarios for how the proposed QDW phase might create multiple superconducting phases for which lower-temperature measurements are required. These scenarios are depicted in Fig. \ref{fig:Phasediagram1}(c). First, if the $T_0$ phase boundary indeed meets at the multicritical point [see Fig. \ref{fig:Phasediagram1}(c)(i)], this would suggest that SC1 is a phase for which SC and the proposed QDW phase below $T_0$ coexist, while SC2 possesses only SC order. Such a scenario would also allow us to infer the order of the field-induced transition within the superconducting state. Since it is thermodynamically forbidden for three second-order phase transition lines to meet at a tricritical point \cite{Yip}, the field-induced transition was thought to be first order \cite{Khim}. But a $T_0$ phase boundary merging at the multicritical point would remove this thermodynamic constraint and allow the field-induced transition to be second ordered. Alternatively, we can envision the $T_0$ phase boundary terminating at SC2 for a field above the multicritical point [see Fig. \ref{fig:Phasediagram1}(c)(ii)]. This would suggest a scenario where the fluctuations associated with a field-induced QDW quantum critical point (QCP) mediate the SC2 phase similar to how SC is often found in the vicinity of an antiferromagnetic QCP \cite{Mathur,Movshovich}. It is worth noting that one might expect evidence for a phase boundary within SC2. Otherwise, this scenario is thermodynamically forbidden \cite{Yip} under the assumption that the proposed QDW transition is second ordered in field. As stated above, lower temperature measurements are needed to distinguish between these two scenarios.

\subsection{Summary}
In summary, we performed anisotropic magnetotransport measurements in the heavy-fermion superconductor CeRh$_2$As$_2$. We find that the proposed quadrupole-density-wave phase at $T_0$ manifests in both $\rho_a (T)$ and $\rho_c (T)$. Furthermore, for a magnetic field applied along the $a$ axis, $T_0$ rises with increasing magnetic fields and a field-induced transition occurs at $\mu_0 H_m \approx $ 8.1~T, where the suggested quadrupole-density- wave phase changes to a new broken symmetry state. In comparison, for a magnetic field applied along the $c$ axis, the quadrupole-density-wave phase at $T_0$ is suppressed below our base temperature $\sim$0.35~K by a field $\mu_0 H \sim$ 4.5~T, close to the transition field between two different SC phases. Our results suggest that the quadrupole-density-wave phase plays a key role in the unique superconducting properties of CeRh$_2$As$_2$ and leads to multiphase superconductivity when a magnetic field is applied along the $c$ direction.

\begin{acknowledgments}
Work at Los Alamos was supported by the U.S. Department of Energy,  Office of Science, National Quantum Information Science Research Centers, Quantum Science Center. Scanning electron microscope imaging and focused ion beam milling was supported by the Center for Integrated Nanotechnologies, an Office of Science User Facility operated for the U.S. Department of Energy Office of Science.  Y.L. was supported by the Los Alamos Laboratory Directed Research and Development program. Sample synthesis by E.D.B. was supported by the U.S. Department of Energy (DOE), Office of Basic Energy Sciences, Division of Materials Science and Engineering under project “Quantum Fluctuations in Narrow-Band Systems." We acknowledge J. D. Thompson and C. Girod for their helpful suggestions in the preparation of this manuscript.
\end{acknowledgments}

\bibliography{CeRh2As2_transport_V4}

\begin{thebibliography}{18}%
\makeatletter
\providecommand \@ifxundefined [1]{%
 \@ifx{#1\undefined}
}%
\providecommand \@ifnum [1]{%
 \ifnum #1\expandafter \@firstoftwo
 \else \expandafter \@secondoftwo
 \fi
}%
\providecommand \@ifx [1]{%
 \ifx #1\expandafter \@firstoftwo
 \else \expandafter \@secondoftwo
 \fi
}%
\providecommand \natexlab [1]{#1}%
\providecommand \enquote  [1]{``#1''}%
\providecommand \bibnamefont  [1]{#1}%
\providecommand \bibfnamefont [1]{#1}%
\providecommand \citenamefont [1]{#1}%
\providecommand \href@noop [0]{\@secondoftwo}%
\providecommand \href [0]{\begingroup \@sanitize@url \@href}%
\providecommand \@href[1]{\@@startlink{#1}\@@href}%
\providecommand \@@href[1]{\endgroup#1\@@endlink}%
\providecommand \@sanitize@url [0]{\catcode `\\12\catcode `\$12\catcode
  `\&12\catcode `\#12\catcode `\^12\catcode `\_12\catcode `\%12\relax}%
\providecommand \@@startlink[1]{}%
\providecommand \@@endlink[0]{}%
\providecommand \url  [0]{\begingroup\@sanitize@url \@url }%
\providecommand \@url [1]{\endgroup\@href {#1}{\urlprefix }}%
\providecommand \urlprefix  [0]{URL }%
\providecommand \Eprint [0]{\href }%
\providecommand \doibase [0]{https://doi.org/}%
\providecommand \selectlanguage [0]{\@gobble}%
\providecommand \bibinfo  [0]{\@secondoftwo}%
\providecommand \bibfield  [0]{\@secondoftwo}%
\providecommand \translation [1]{[#1]}%
\providecommand \BibitemOpen [0]{}%
\providecommand \bibitemStop [0]{}%
\providecommand \bibitemNoStop [0]{.\EOS\space}%
\providecommand \EOS [0]{\spacefactor3000\relax}%
\providecommand \BibitemShut  [1]{\csname bibitem#1\endcsname}%
\let\auto@bib@innerbib\@empty
%</preamble>
\bibitem [{\citenamefont {Joynt}\ and\ \citenamefont
  {Taillefer}(2002)}]{Joynt}%
  \BibitemOpen
  \bibfield  {author} {\bibinfo {author} {\bibfnamefont {R.}~\bibnamefont
  {Joynt}}\ and\ \bibinfo {author} {\bibfnamefont {L.}~\bibnamefont
  {Taillefer}},\ }\bibfield  {title} {\bibinfo {title} {The superconducting
  phases of {UP}t$_3$},\ }\href {https://doi.org/10.1103/RevModPhys.74.235}
  {\bibfield  {journal} {\bibinfo  {journal} {Rev. Mod. Phys.}\ }\textbf
  {\bibinfo {volume} {74}},\ \bibinfo {pages} {235} (\bibinfo {year}
  {2002})}\BibitemShut {NoStop}%
\bibitem [{\citenamefont {Heffner}\ \emph {et~al.}(1990)\citenamefont
  {Heffner}, \citenamefont {Smith}, \citenamefont {Willis}, \citenamefont
  {Birrer}, \citenamefont {Baines}, \citenamefont {Gygax}, \citenamefont
  {Hitti}, \citenamefont {Lippelt}, \citenamefont {Ott}, \citenamefont
  {Schenck}, \citenamefont {Knetsch}, \citenamefont {Mydosh},\ and\
  \citenamefont {MacLaughlin}}]{Heffner}%
  \BibitemOpen
  \bibfield  {author} {\bibinfo {author} {\bibfnamefont {R.~H.}\ \bibnamefont
  {Heffner}}, \bibinfo {author} {\bibfnamefont {J.~L.}\ \bibnamefont {Smith}},
  \bibinfo {author} {\bibfnamefont {J.~O.}\ \bibnamefont {Willis}}, \bibinfo
  {author} {\bibfnamefont {P.}~\bibnamefont {Birrer}}, \bibinfo {author}
  {\bibfnamefont {C.}~\bibnamefont {Baines}}, \bibinfo {author} {\bibfnamefont
  {F.~N.}\ \bibnamefont {Gygax}}, \bibinfo {author} {\bibfnamefont
  {B.}~\bibnamefont {Hitti}}, \bibinfo {author} {\bibfnamefont
  {E.}~\bibnamefont {Lippelt}}, \bibinfo {author} {\bibfnamefont {H.~R.}\
  \bibnamefont {Ott}}, \bibinfo {author} {\bibfnamefont {A.}~\bibnamefont
  {Schenck}}, \bibinfo {author} {\bibfnamefont {E.~A.}\ \bibnamefont
  {Knetsch}}, \bibinfo {author} {\bibfnamefont {J.~A.}\ \bibnamefont
  {Mydosh}},\ and\ \bibinfo {author} {\bibfnamefont {D.~E.}\ \bibnamefont
  {MacLaughlin}},\ }\bibfield  {title} {\bibinfo {title} {New phase diagram for
  ({U,Th}){B}e$_{13}$: A muon-spin-resonance and {H}$_{c1}$ study},\ }\href
  {https://doi.org/10.1103/PhysRevLett.65.2816} {\bibfield  {journal} {\bibinfo
   {journal} {Phys. Rev. Lett.}\ }\textbf {\bibinfo {volume} {65}},\ \bibinfo
  {pages} {2816} (\bibinfo {year} {1990})}\BibitemShut {NoStop}%
\bibitem [{\citenamefont {Bauer}\ \emph {et~al.}(2002)\citenamefont {Bauer},
  \citenamefont {Frederick}, \citenamefont {Ho}, \citenamefont {Zapf},\ and\
  \citenamefont {Maple}}]{Bauer}%
  \BibitemOpen
  \bibfield  {author} {\bibinfo {author} {\bibfnamefont {E.~D.}\ \bibnamefont
  {Bauer}}, \bibinfo {author} {\bibfnamefont {N.~A.}\ \bibnamefont
  {Frederick}}, \bibinfo {author} {\bibfnamefont {P.-C.}\ \bibnamefont {Ho}},
  \bibinfo {author} {\bibfnamefont {V.~S.}\ \bibnamefont {Zapf}},\ and\
  \bibinfo {author} {\bibfnamefont {M.~B.}\ \bibnamefont {Maple}},\ }\bibfield
  {title} {\bibinfo {title} {Superconductivity and heavy fermion behavior in
  {P}r{O}s$_{4}${S}b$_{12}$},\ }\href
  {https://doi.org/10.1103/PhysRevB.65.100506} {\bibfield  {journal} {\bibinfo
  {journal} {Phys. Rev. B}\ }\textbf {\bibinfo {volume} {65}},\ \bibinfo
  {pages} {100506(R)} (\bibinfo {year} {2002})}\BibitemShut {NoStop}%
\bibitem [{\citenamefont {Izawa}\ \emph {et~al.}(2003)\citenamefont {Izawa},
  \citenamefont {Nakajima}, \citenamefont {Goryo}, \citenamefont {Matsuda},
  \citenamefont {Osaki}, \citenamefont {Sugawara}, \citenamefont {Sato},
  \citenamefont {Thalmeier},\ and\ \citenamefont {Maki}}]{Izawa}%
  \BibitemOpen
  \bibfield  {author} {\bibinfo {author} {\bibfnamefont {K.}~\bibnamefont
  {Izawa}}, \bibinfo {author} {\bibfnamefont {Y.}~\bibnamefont {Nakajima}},
  \bibinfo {author} {\bibfnamefont {J.}~\bibnamefont {Goryo}}, \bibinfo
  {author} {\bibfnamefont {Y.}~\bibnamefont {Matsuda}}, \bibinfo {author}
  {\bibfnamefont {S.}~\bibnamefont {Osaki}}, \bibinfo {author} {\bibfnamefont
  {H.}~\bibnamefont {Sugawara}}, \bibinfo {author} {\bibfnamefont
  {H.}~\bibnamefont {Sato}}, \bibinfo {author} {\bibfnamefont {P.}~\bibnamefont
  {Thalmeier}},\ and\ \bibinfo {author} {\bibfnamefont {K.}~\bibnamefont
  {Maki}},\ }\bibfield  {title} {\bibinfo {title} {Multiple superconducting
  phases in new heavy fermion superconductor {P}r{O}s$_{4}${S}b$_{12}$},\
  }\href {https://doi.org/10.1103/PhysRevLett.90.117001} {\bibfield  {journal}
  {\bibinfo  {journal} {Phys. Rev. Lett.}\ }\textbf {\bibinfo {volume} {90}},\
  \bibinfo {pages} {117001} (\bibinfo {year} {2003})}\BibitemShut {NoStop}%
\bibitem [{\citenamefont {B.~Maple}\ \emph {et~al.}(2002)\citenamefont
  {B.~Maple}, \citenamefont {Ho}, \citenamefont {S.~Zapf}, \citenamefont
  {A.~Frederick}, \citenamefont {D.~Bauer}, \citenamefont {M.~Yuhasz},
  \citenamefont {M.~Woodward},\ and\ \citenamefont {W.~Lynn}}]{Maple}%
  \BibitemOpen
  \bibfield  {author} {\bibinfo {author} {\bibfnamefont {M.}~\bibnamefont
  {B.~Maple}}, \bibinfo {author} {\bibfnamefont {P.-C.}\ \bibnamefont {Ho}},
  \bibinfo {author} {\bibfnamefont {V.}~\bibnamefont {S.~Zapf}}, \bibinfo
  {author} {\bibfnamefont {N.}~\bibnamefont {A.~Frederick}}, \bibinfo {author}
  {\bibfnamefont {E.}~\bibnamefont {D.~Bauer}}, \bibinfo {author}
  {\bibfnamefont {W.}~\bibnamefont {M.~Yuhasz}}, \bibinfo {author}
  {\bibfnamefont {F.}~\bibnamefont {M.~Woodward}},\ and\ \bibinfo {author}
  {\bibfnamefont {J.}~\bibnamefont {W.~Lynn}},\ }\bibfield  {title} {\bibinfo
  {title} {Heavy fermion superconductivity in the filled skutterudite compound
  {P}r{O}s$_{4}${S}b$_{12}$},\ }\href {https://doi.org/10.1143/JPSJS.71S.23}
  {\bibfield  {journal} {\bibinfo  {journal} {Journal of the Physical Society
  of Japan}\ }\textbf {\bibinfo {volume} {71}},\ \bibinfo {pages} {23}
  (\bibinfo {year} {2002})}\BibitemShut {NoStop}%
\bibitem [{\citenamefont {Braithwaite}\ \emph {et~al.}(2019)\citenamefont
  {Braithwaite}, \citenamefont {Vališka}, \citenamefont {Knebel},
  \citenamefont {Lapertot}, \citenamefont {Brison}, \citenamefont {Pourret},
  \citenamefont {Zhitomirsky}, \citenamefont {Flouquet}, \citenamefont
  {Honda},\ and\ \citenamefont {Aoki}}]{Braithwaite2019}%
  \BibitemOpen
  \bibfield  {author} {\bibinfo {author} {\bibfnamefont {D.}~\bibnamefont
  {Braithwaite}}, \bibinfo {author} {\bibfnamefont {M.}~\bibnamefont
  {Vališka}}, \bibinfo {author} {\bibfnamefont {G.}~\bibnamefont {Knebel}},
  \bibinfo {author} {\bibfnamefont {G.}~\bibnamefont {Lapertot}}, \bibinfo
  {author} {\bibfnamefont {J.-P.}\ \bibnamefont {Brison}}, \bibinfo {author}
  {\bibfnamefont {A.}~\bibnamefont {Pourret}}, \bibinfo {author} {\bibfnamefont
  {M.~E.}\ \bibnamefont {Zhitomirsky}}, \bibinfo {author} {\bibfnamefont
  {J.}~\bibnamefont {Flouquet}}, \bibinfo {author} {\bibfnamefont
  {F.}~\bibnamefont {Honda}},\ and\ \bibinfo {author} {\bibfnamefont
  {D.}~\bibnamefont {Aoki}},\ }\bibfield  {title} {\bibinfo {title} {Multiple
  superconducting phases in a nearly ferromagnetic system},\ }\href
  {https://doi.org/10.1038/s42005-019-0248-z} {\bibfield  {journal} {\bibinfo
  {journal} {Commun Phys}\ }\textbf {\bibinfo {volume} {2}},\ \bibinfo {pages}
  {147} (\bibinfo {year} {2019})}\BibitemShut {NoStop}%
\bibitem [{\citenamefont {Aoki}\ \emph {et~al.}(2020)\citenamefont {Aoki},
  \citenamefont {Honda}, \citenamefont {Knebel}, \citenamefont {Braithwaite},
  \citenamefont {Nakamura}, \citenamefont {Li}, \citenamefont {Homma},
  \citenamefont {Shimizu}, \citenamefont {Sato}, \citenamefont {Brison},\ and\
  \citenamefont {Flouquet}}]{Aoki2020}%
  \BibitemOpen
  \bibfield  {author} {\bibinfo {author} {\bibfnamefont {D.}~\bibnamefont
  {Aoki}}, \bibinfo {author} {\bibfnamefont {F.}~\bibnamefont {Honda}},
  \bibinfo {author} {\bibfnamefont {G.}~\bibnamefont {Knebel}}, \bibinfo
  {author} {\bibfnamefont {D.}~\bibnamefont {Braithwaite}}, \bibinfo {author}
  {\bibfnamefont {A.}~\bibnamefont {Nakamura}}, \bibinfo {author}
  {\bibfnamefont {D.}~\bibnamefont {Li}}, \bibinfo {author} {\bibfnamefont
  {Y.}~\bibnamefont {Homma}}, \bibinfo {author} {\bibfnamefont
  {Y.}~\bibnamefont {Shimizu}}, \bibinfo {author} {\bibfnamefont {Y.~J.}\
  \bibnamefont {Sato}}, \bibinfo {author} {\bibfnamefont {J.-P.}\ \bibnamefont
  {Brison}},\ and\ \bibinfo {author} {\bibfnamefont {J.}~\bibnamefont
  {Flouquet}},\ }\bibfield  {title} {\bibinfo {title} {Multiple superconducting
  phases and unusual enhancement of the upper critical field in {UT}e$_2$},\
  }\href {https://doi.org/10.7566/jpsj.89.053705} {\bibfield  {journal}
  {\bibinfo  {journal} {J. Phys. Soc. Jpn.}\ }\textbf {\bibinfo {volume}
  {89}},\ \bibinfo {pages} {053705} (\bibinfo {year} {2020})}\BibitemShut
  {NoStop}%
\bibitem [{\citenamefont {Thomas}\ \emph {et~al.}(2020)\citenamefont {Thomas},
  \citenamefont {Santos}, \citenamefont {Christensen}, \citenamefont {Asaba},
  \citenamefont {Ronning}, \citenamefont {Thompson}, \citenamefont {Bauer},
  \citenamefont {Fernandes}, \citenamefont {Fabbris},\ and\ \citenamefont
  {Rosa}}]{Thomas}%
  \BibitemOpen
  \bibfield  {author} {\bibinfo {author} {\bibfnamefont {S.~M.}\ \bibnamefont
  {Thomas}}, \bibinfo {author} {\bibfnamefont {F.~B.}\ \bibnamefont {Santos}},
  \bibinfo {author} {\bibfnamefont {M.~H.}\ \bibnamefont {Christensen}},
  \bibinfo {author} {\bibfnamefont {T.}~\bibnamefont {Asaba}}, \bibinfo
  {author} {\bibfnamefont {F.}~\bibnamefont {Ronning}}, \bibinfo {author}
  {\bibfnamefont {J.~D.}\ \bibnamefont {Thompson}}, \bibinfo {author}
  {\bibfnamefont {E.~D.}\ \bibnamefont {Bauer}}, \bibinfo {author}
  {\bibfnamefont {R.~M.}\ \bibnamefont {Fernandes}}, \bibinfo {author}
  {\bibfnamefont {G.}~\bibnamefont {Fabbris}},\ and\ \bibinfo {author}
  {\bibfnamefont {P.~F.~S.}\ \bibnamefont {Rosa}},\ }\bibfield  {title}
  {\bibinfo {title} {Evidence for a pressure-induced antiferromagnetic quantum
  critical point in intermediate-valence {UT}e$_2$},\ }\href
  {https://doi.org/10.1126/sciadv.abc8709} {\bibfield  {journal} {\bibinfo
  {journal} {Science Advances}\ }\textbf {\bibinfo {volume} {6}},\ \bibinfo
  {pages} {eabc8709} (\bibinfo {year} {2020})}\BibitemShut {NoStop}%
\bibitem [{\citenamefont {Khim}\ \emph {et~al.}(2021)\citenamefont {Khim},
  \citenamefont {Landaeta}, \citenamefont {Banda}, \citenamefont {Bannor},
  \citenamefont {Brando}, \citenamefont {Brydon}, \citenamefont {Hafner},
  \citenamefont {Küchler}, \citenamefont {Cardoso-Gil}, \citenamefont
  {Stockert}, \citenamefont {Mackenzie}, \citenamefont {Agterberg},
  \citenamefont {Geibel},\ and\ \citenamefont {Hassinger}}]{Khim}%
  \BibitemOpen
  \bibfield  {author} {\bibinfo {author} {\bibfnamefont {S.}~\bibnamefont
  {Khim}}, \bibinfo {author} {\bibfnamefont {J.~F.}\ \bibnamefont {Landaeta}},
  \bibinfo {author} {\bibfnamefont {J.}~\bibnamefont {Banda}}, \bibinfo
  {author} {\bibfnamefont {N.}~\bibnamefont {Bannor}}, \bibinfo {author}
  {\bibfnamefont {M.}~\bibnamefont {Brando}}, \bibinfo {author} {\bibfnamefont
  {P.~M.~R.}\ \bibnamefont {Brydon}}, \bibinfo {author} {\bibfnamefont
  {D.}~\bibnamefont {Hafner}}, \bibinfo {author} {\bibfnamefont
  {R.}~\bibnamefont {Küchler}}, \bibinfo {author} {\bibfnamefont
  {R.}~\bibnamefont {Cardoso-Gil}}, \bibinfo {author} {\bibfnamefont
  {U.}~\bibnamefont {Stockert}}, \bibinfo {author} {\bibfnamefont {A.~P.}\
  \bibnamefont {Mackenzie}}, \bibinfo {author} {\bibfnamefont {D.~F.}\
  \bibnamefont {Agterberg}}, \bibinfo {author} {\bibfnamefont {C.}~\bibnamefont
  {Geibel}},\ and\ \bibinfo {author} {\bibfnamefont {E.}~\bibnamefont
  {Hassinger}},\ }\bibfield  {title} {\bibinfo {title} {Field-induced
  transition within the superconducting state of {C}e{R}h$_2${A}s$_2$},\ }\href
  {https://doi.org/10.1126/science.abe7518} {\bibfield  {journal} {\bibinfo
  {journal} {Science}\ }\textbf {\bibinfo {volume} {373}},\ \bibinfo {pages}
  {1012} (\bibinfo {year} {2021})}\BibitemShut {NoStop}%
\bibitem [{\citenamefont {{E}l {G}hadraoui}\ \emph {et~al.}(1988)\citenamefont
  {{E}l {G}hadraoui}, \citenamefont {Pivan},\ and\ \citenamefont
  {Guérin}}]{Madar1987}%
  \BibitemOpen
  \bibfield  {author} {\bibinfo {author} {\bibfnamefont {E.}~\bibnamefont {{E}l
  {G}hadraoui}}, \bibinfo {author} {\bibfnamefont {J.}~\bibnamefont {Pivan}},\
  and\ \bibinfo {author} {\bibfnamefont {R.}~\bibnamefont {Guérin}},\
  }\bibfield  {title} {\bibinfo {title} {New ternary pnictides
  {MN}i$_{0.75}$x$_2$ ({M} = {Z}r, {H}f) with a defective
  {C}a{B}e$_2${G}e$_2$-type structure-structure and properties},\ }\href
  {https://www.sciencedirect.com/science/article/pii/0022508888904341}
  {\bibfield  {journal} {\bibinfo  {journal} {Journal of the Less Common
  Metals}\ }\textbf {\bibinfo {volume} {136}},\ \bibinfo {pages} {303}
  (\bibinfo {year} {1988})}\BibitemShut {NoStop}%
\bibitem [{\citenamefont {Landaeta}\ \emph {et~al.}(2022)\citenamefont
  {Landaeta}, \citenamefont {Khanenko}, \citenamefont {Cavanagh}, \citenamefont
  {Geibel}, \citenamefont {Khim}, \citenamefont {Mishra}, \citenamefont
  {Sheikin}, \citenamefont {Brydon}, \citenamefont {Agterberg}, \citenamefont
  {Brando},\ and\ \citenamefont {Hassinger}}]{Landaeta}%
  \BibitemOpen
  \bibfield  {author} {\bibinfo {author} {\bibfnamefont {J.~F.}\ \bibnamefont
  {Landaeta}}, \bibinfo {author} {\bibfnamefont {P.}~\bibnamefont {Khanenko}},
  \bibinfo {author} {\bibfnamefont {D.~C.}\ \bibnamefont {Cavanagh}}, \bibinfo
  {author} {\bibfnamefont {C.}~\bibnamefont {Geibel}}, \bibinfo {author}
  {\bibfnamefont {S.}~\bibnamefont {Khim}}, \bibinfo {author} {\bibfnamefont
  {S.}~\bibnamefont {Mishra}}, \bibinfo {author} {\bibfnamefont
  {I.}~\bibnamefont {Sheikin}}, \bibinfo {author} {\bibfnamefont {P.~M.~R.}\
  \bibnamefont {Brydon}}, \bibinfo {author} {\bibfnamefont {D.~F.}\
  \bibnamefont {Agterberg}}, \bibinfo {author} {\bibfnamefont {M.}~\bibnamefont
  {Brando}},\ and\ \bibinfo {author} {\bibfnamefont {E.}~\bibnamefont
  {Hassinger}},\ }\bibfield  {title} {\bibinfo {title} {Field-angle dependence
  reveals odd-parity superconductivity in {C}e{R}h$_2${A}s$_2$},\ }\href
  {https://doi.org/10.1103/PhysRevX.12.031001} {\bibfield  {journal} {\bibinfo
  {journal} {Phys. Rev. X}\ }\textbf {\bibinfo {volume} {12}},\ \bibinfo
  {pages} {031001} (\bibinfo {year} {2022})}\BibitemShut {NoStop}%
\bibitem [{\citenamefont {Kibune}\ \emph {et~al.}(2022)\citenamefont {Kibune},
  \citenamefont {Kitagawa}, \citenamefont {Kinjo}, \citenamefont {Ogata},
  \citenamefont {Manago}, \citenamefont {Taniguchi}, \citenamefont {Ishida},
  \citenamefont {Brando}, \citenamefont {Hassinger}, \citenamefont {Rosner},
  \citenamefont {Geibel},\ and\ \citenamefont {Khim}}]{Kibune}%
  \BibitemOpen
  \bibfield  {author} {\bibinfo {author} {\bibfnamefont {M.}~\bibnamefont
  {Kibune}}, \bibinfo {author} {\bibfnamefont {S.}~\bibnamefont {Kitagawa}},
  \bibinfo {author} {\bibfnamefont {K.}~\bibnamefont {Kinjo}}, \bibinfo
  {author} {\bibfnamefont {S.}~\bibnamefont {Ogata}}, \bibinfo {author}
  {\bibfnamefont {M.}~\bibnamefont {Manago}}, \bibinfo {author} {\bibfnamefont
  {T.}~\bibnamefont {Taniguchi}}, \bibinfo {author} {\bibfnamefont
  {K.}~\bibnamefont {Ishida}}, \bibinfo {author} {\bibfnamefont
  {M.}~\bibnamefont {Brando}}, \bibinfo {author} {\bibfnamefont
  {E.}~\bibnamefont {Hassinger}}, \bibinfo {author} {\bibfnamefont
  {H.}~\bibnamefont {Rosner}}, \bibinfo {author} {\bibfnamefont
  {C.}~\bibnamefont {Geibel}},\ and\ \bibinfo {author} {\bibfnamefont
  {S.}~\bibnamefont {Khim}},\ }\bibfield  {title} {\bibinfo {title}
  {Observation of antiferromagnetic order as odd-parity multipoles inside the
  superconducting phase in {C}e{R}h$_2${A}s$_2$},\ }\href
  {https://doi.org/10.1103/PhysRevLett.128.057002} {\bibfield  {journal}
  {\bibinfo  {journal} {Phys. Rev. Lett.}\ }\textbf {\bibinfo {volume} {128}},\
  \bibinfo {pages} {057002} (\bibinfo {year} {2022})}\BibitemShut {NoStop}%
\bibitem [{\citenamefont {Hafner}\ \emph {et~al.}(2022)\citenamefont {Hafner},
  \citenamefont {Khanenko}, \citenamefont {Eljaouhari}, \citenamefont
  {K\"uchler}, \citenamefont {Banda}, \citenamefont {Bannor}, \citenamefont
  {L\"uhmann}, \citenamefont {Landaeta}, \citenamefont {Mishra}, \citenamefont
  {Sheikin}, \citenamefont {Hassinger}, \citenamefont {Khim}, \citenamefont
  {Geibel}, \citenamefont {Zwicknagl},\ and\ \citenamefont {Brando}}]{Hafner}%
  \BibitemOpen
  \bibfield  {author} {\bibinfo {author} {\bibfnamefont {D.}~\bibnamefont
  {Hafner}}, \bibinfo {author} {\bibfnamefont {P.}~\bibnamefont {Khanenko}},
  \bibinfo {author} {\bibfnamefont {E.-O.}\ \bibnamefont {Eljaouhari}},
  \bibinfo {author} {\bibfnamefont {R.}~\bibnamefont {K\"uchler}}, \bibinfo
  {author} {\bibfnamefont {J.}~\bibnamefont {Banda}}, \bibinfo {author}
  {\bibfnamefont {N.}~\bibnamefont {Bannor}}, \bibinfo {author} {\bibfnamefont
  {T.}~\bibnamefont {L\"uhmann}}, \bibinfo {author} {\bibfnamefont {J.~F.}\
  \bibnamefont {Landaeta}}, \bibinfo {author} {\bibfnamefont {S.}~\bibnamefont
  {Mishra}}, \bibinfo {author} {\bibfnamefont {I.}~\bibnamefont {Sheikin}},
  \bibinfo {author} {\bibfnamefont {E.}~\bibnamefont {Hassinger}}, \bibinfo
  {author} {\bibfnamefont {S.}~\bibnamefont {Khim}}, \bibinfo {author}
  {\bibfnamefont {C.}~\bibnamefont {Geibel}}, \bibinfo {author} {\bibfnamefont
  {G.}~\bibnamefont {Zwicknagl}},\ and\ \bibinfo {author} {\bibfnamefont
  {M.}~\bibnamefont {Brando}},\ }\bibfield  {title} {\bibinfo {title} {Possible
  quadrupole density wave in the superconducting kondo lattice
  {C}e{R}h$_2${A}s$_2$},\ }\href {https://doi.org/10.1103/PhysRevX.12.011023}
  {\bibfield  {journal} {\bibinfo  {journal} {Phys. Rev. X}\ }\textbf {\bibinfo
  {volume} {12}},\ \bibinfo {pages} {011023} (\bibinfo {year}
  {2022})}\BibitemShut {NoStop}%
\bibitem [{\citenamefont {Ptok}\ \emph {et~al.}(2021)\citenamefont {Ptok},
  \citenamefont {Kapcia}, \citenamefont {Jochym}, \citenamefont
  {\L{}a\ifmmode~\dot{z}\else \.{z}\fi{}ewski}, \citenamefont
  {Ole\ifmmode~\acute{s}\else \'{s}\fi{}},\ and\ \citenamefont
  {Piekarz}}]{Ptok}%
  \BibitemOpen
  \bibfield  {author} {\bibinfo {author} {\bibfnamefont {A.}~\bibnamefont
  {Ptok}}, \bibinfo {author} {\bibfnamefont {K.~J.}\ \bibnamefont {Kapcia}},
  \bibinfo {author} {\bibfnamefont {P.~T.}\ \bibnamefont {Jochym}}, \bibinfo
  {author} {\bibfnamefont {J.}~\bibnamefont {\L{}a\ifmmode~\dot{z}\else
  \.{z}\fi{}ewski}}, \bibinfo {author} {\bibfnamefont {A.~M.}\ \bibnamefont
  {Ole\ifmmode~\acute{s}\else \'{s}\fi{}}},\ and\ \bibinfo {author}
  {\bibfnamefont {P.}~\bibnamefont {Piekarz}},\ }\bibfield  {title} {\bibinfo
  {title} {Electronic and dynamical properties of {C}e{R}h$_2${A}s$_2$: Role of
  {R}h$_2${A}s$_2$ layers and expected orbital order},\ }\href
  {https://doi.org/10.1103/PhysRevB.104.L041109} {\bibfield  {journal}
  {\bibinfo  {journal} {Phys. Rev. B}\ }\textbf {\bibinfo {volume} {104}},\
  \bibinfo {pages} {L041109} (\bibinfo {year} {2021})}\BibitemShut {NoStop}%
\bibitem [{\citenamefont {Cavanagh}\ \emph {et~al.}(2022)\citenamefont
  {Cavanagh}, \citenamefont {Shishidou}, \citenamefont {Weinert}, \citenamefont
  {Brydon},\ and\ \citenamefont {Agterberg}}]{Cavanagh}%
  \BibitemOpen
  \bibfield  {author} {\bibinfo {author} {\bibfnamefont {D.~C.}\ \bibnamefont
  {Cavanagh}}, \bibinfo {author} {\bibfnamefont {T.}~\bibnamefont {Shishidou}},
  \bibinfo {author} {\bibfnamefont {M.}~\bibnamefont {Weinert}}, \bibinfo
  {author} {\bibfnamefont {P.~M.~R.}\ \bibnamefont {Brydon}},\ and\ \bibinfo
  {author} {\bibfnamefont {D.~F.}\ \bibnamefont {Agterberg}},\ }\bibfield
  {title} {\bibinfo {title} {Nonsymmorphic symmetry and field-driven odd-parity
  pairing in {C}e{R}h$_2${A}s$_2$},\ }\href
  {https://doi.org/10.1103/PhysRevB.105.L020505} {\bibfield  {journal}
  {\bibinfo  {journal} {Phys. Rev. B}\ }\textbf {\bibinfo {volume} {105}},\
  \bibinfo {pages} {L020505} (\bibinfo {year} {2022})}\BibitemShut {NoStop}%
\bibitem [{\citenamefont {Yip}\ \emph {et~al.}(1991)\citenamefont {Yip},
  \citenamefont {Li},\ and\ \citenamefont {Kumar}}]{Yip}%
  \BibitemOpen
  \bibfield  {author} {\bibinfo {author} {\bibfnamefont {S.~K.}\ \bibnamefont
  {Yip}}, \bibinfo {author} {\bibfnamefont {T.}~\bibnamefont {Li}},\ and\
  \bibinfo {author} {\bibfnamefont {P.}~\bibnamefont {Kumar}},\ }\bibfield
  {title} {\bibinfo {title} {Thermodynamic considerations and the phase diagram
  of superconducting {UP}t$_3$},\ }\href
  {https://doi.org/10.1103/PhysRevB.43.2742} {\bibfield  {journal} {\bibinfo
  {journal} {Phys. Rev. B}\ }\textbf {\bibinfo {volume} {43}},\ \bibinfo
  {pages} {2742} (\bibinfo {year} {1991})}\BibitemShut {NoStop}%
\bibitem [{\citenamefont {Mathur}\ \emph {et~al.}(1998)\citenamefont {Mathur},
  \citenamefont {Grosche}, \citenamefont {Julian}, \citenamefont {Walker},
  \citenamefont {Freye}, \citenamefont {Haselwimmer},\ and\ \citenamefont
  {Lonzarich}}]{Mathur}%
  \BibitemOpen
  \bibfield  {author} {\bibinfo {author} {\bibfnamefont {N.~D.}\ \bibnamefont
  {Mathur}}, \bibinfo {author} {\bibfnamefont {F.~M.}\ \bibnamefont {Grosche}},
  \bibinfo {author} {\bibfnamefont {S.~R.}\ \bibnamefont {Julian}}, \bibinfo
  {author} {\bibfnamefont {I.~R.}\ \bibnamefont {Walker}}, \bibinfo {author}
  {\bibfnamefont {D.~M.}\ \bibnamefont {Freye}}, \bibinfo {author}
  {\bibfnamefont {R.~K.~W.}\ \bibnamefont {Haselwimmer}},\ and\ \bibinfo
  {author} {\bibfnamefont {G.~G.}\ \bibnamefont {Lonzarich}},\ }\bibfield
  {title} {\bibinfo {title} {{Magnetically mediated superconductivity in heavy
  fermion compounds}},\ }\href {https://doi.org/10.1038/27838} {\bibfield
  {journal} {\bibinfo  {journal} {Nature}\ }\textbf {\bibinfo {volume} {394}},\
  \bibinfo {pages} {39} (\bibinfo {year} {1998})}\BibitemShut {NoStop}%
\bibitem [{\citenamefont {Movshovich}\ \emph {et~al.}(1996)\citenamefont
  {Movshovich}, \citenamefont {Graf}, \citenamefont {Mandrus}, \citenamefont
  {Thompson}, \citenamefont {Smith},\ and\ \citenamefont {Fisk}}]{Movshovich}%
  \BibitemOpen
  \bibfield  {author} {\bibinfo {author} {\bibfnamefont {R.}~\bibnamefont
  {Movshovich}}, \bibinfo {author} {\bibfnamefont {T.}~\bibnamefont {Graf}},
  \bibinfo {author} {\bibfnamefont {D.}~\bibnamefont {Mandrus}}, \bibinfo
  {author} {\bibfnamefont {J.~D.}\ \bibnamefont {Thompson}}, \bibinfo {author}
  {\bibfnamefont {J.~L.}\ \bibnamefont {Smith}},\ and\ \bibinfo {author}
  {\bibfnamefont {Z.}~\bibnamefont {Fisk}},\ }\bibfield  {title} {\bibinfo
  {title} {Superconductivity in heavy-fermion {C}e{R}h$_2${S}i$_2$},\ }\href
  {https://doi.org/10.1103/PhysRevB.53.8241} {\bibfield  {journal} {\bibinfo
  {journal} {Phys. Rev. B}\ }\textbf {\bibinfo {volume} {53}},\ \bibinfo
  {pages} {8241} (\bibinfo {year} {1996})}\BibitemShut {NoStop}%
\end{thebibliography}%
\end{document}